# Physics Origin of Universal Unusual Magnetoresistance

Lijun Zhu[1,2*], Qianbiao Liu[1], and Xiangrong Wang[3,4*]

1. State Key Laboratory of Superlattices and Microstructures, Institute of Semiconductors, Chinese Academy of Sciences, Beijing 100083, China
2. College of Materials Science and Opto-Electronic Technology, University of Chinese Academy of Sciences, Beijing 100049, China
3. School of Science and Engineering, Chinese University of Hong Kong, Shenzhen, Shenzhen 51817, China
4. Physics Department, The Hong Kong University of Science and Technology, Clear Water Bay, Kowloon, Hong Kong
*ljzhu@semi.ac.cn; phxwan@ cuhk.edu.cn

The discovery of the unusual magnetoresistance (UMR) during the rotation of magnetization in the plane perpendicular to the electric current, which has been typically attributed to magnetization-dependent interfacial reflection of spin current, has brought remarkable impacts on the understanding and application of a variety of spintronic phenomena. Here, we report that giant UMR occurs also in single-layer magnetic metals and exhibits high-order contributions and a universal sum rule, which agree well with the physics origin of the recently proposed two-vector magnetoresistance that simply considers electron scattering by the magnetization vector and interfacial electric field, without the need for any relevance to spin current. Revisiting of the literature data reveals that the most representative data that were used to claim spin Hall magnetoresistance or other magnetoresistances related or unrelated to spin current can be understood unifiedly by the two-vector MR theory. Experimental and theoretical results against spin-current-related magnetoresistances, but not the two-vector magnetoresistance, are discussed.

A dramatic discovery in spintronics is the unusual magnetoresistance (UMR) that the longitudinal resistivity ($\rho$) of a heavy metal (HM) in contact with a magnetic insulator (e.g., YIG = $Y_3Fe_5O_{12}$) varies with the rotation of magnetization in the plane perpendicular to the electric current [1-6]. This newly observed UMR stimulated the development of the spin Hall magnetoresistance (SMR) theory [1-4], in which the absorption/reflection of spin angular momentum at the magnetic interface was proposed to cause a $\cos^2\beta$ dependent resistivity variation, with $\beta$ being the unit vector of the magnetization ($\vec{m}$) relative to the sample normal. So far, the SMR theory has been used to interpret UMR and its transverse counterpart (the planar Hall effect) in bilayers of a HM and a magnetic layer (either insulating or metallic) in a variety of experiments, e.g., magnetoresistance [6-12], spin-torque ferromagnetic resonance [13,14], harmonic Hall voltage [15], magnetic field sensing [16], and magnetization or Néel-vector switching [17-21]. However, the SMR theory is questioned in quantifying the spin Hall ratios of spin-current generators [15,22-24] and in accounting for the strong MR in magnetic systems with no strong spin Hall effect [25]. Thus, alternative spin-current-related MR models were also proposed in the literature to explain the "SMR-like" MR (e.g., Rashba-Edelstein MR [23,24], spin-orbit MR [26], anomalous Hall MR [27], orbital Hall magnetoresistance [28], orbital Rashba-Edelstein MR [29], Hanle MR [30]).

In contrast, a recent symmetry-analysis theory [31] has proposed a two-vector model that the UMR arises simply from electron scattering by the magnetization (with the macroscopic vector $\vec{m}$) and interfacial electric field (with the macroscopic vector of the surface normal $\vec{n}$). The two-vector UMR has three characteristics: i) universal occurrence at the interface of any magnetic layer and without any relevance to spin current or spin polarization; ii) potential presence of high-order contributions ($n \geq 2$) in addition to the first-order contribution ($n = 1$), i.e.,

$$\rho = \rho_0 + \sum_n \Delta\rho_{n\theta}\cos^{2n}\theta, \qquad (1)$$

where $\theta$ represents the angle of the magnetization in the $xy$ ($\alpha$), $yz$ ($\beta$), $zx$ ($\gamma$) planes relative to the $x$, $z$, $z$ directions (see Fig. 1(a) for the coordinates), respectively, $\rho_0 = \rho_y$ for the $\alpha$ and $\beta$ scans and $\rho_0 = \rho_z$ in the $\gamma$ scan ($\rho_z$ is the resistivity when $\vec{m}/\!/\vec{n}$), $\Delta\rho_{n\alpha}$, $\Delta\rho_{n\beta}$, and $\Delta\rho_{n\gamma}$ are the magnitudes of the $n$th-order MR contributions in the $\alpha$, $\beta$, and $\gamma$ scans; (iii) universal sum rule of the MR contributions, i.e.,

$$\sum_n \Delta\rho_{n\alpha} + \sum_n \Delta\rho_{n\gamma} = \sum_n \Delta\rho_{n\beta} \qquad (2)$$

The two-vector MR theory, if correct, would mean that the cos2$\beta$ dependent UMR is a rather universal effect (despite the different magnitudes in different systems) and cannot be taken as the signature of the spin-current MR models. Given the fundamental and widespread impact of the UMR effect [1-30], the experimental test of the two-vector MR theory and a unified, precise understanding of the physics origin of the UMR are urgently required.

In this letter, we report that the giant UMR can occur in single-layer magnetic metals and exhibits all the characteristics of the two-vector UMR. Revisiting the literature data reveals that the most representative data that were used to claim SMR and other spin-current-related MRs can be understood unifiedly by the two-vector MR theory, without the need for any relevance to spin current.

For this work, we first sputter-deposited CoPt (=$Co_{0.5}Pt_{0.5}$) single layers with different thicknesses ($t_{CoPt}$) of 4, 8, 12, 16, and 24 nm at room temperature on thermally oxidized Si substrates. Each sample is protected subsequently by a 2 nm MgO layer (we note as $SiO_2$/CoPt $t_{CoPt}$/MgO) and a 1.5 nm Ta layer that was fully oxidized upon exposure to the atmosphere. We also prepare two samples with symmetric interfaces, Si/$SiO_2$/MgO 2/CoPt 16/MgO 2/Ta 1.5 (noted as MgO/CoPt 16/MgO) and Si/$SiO_2$/Hf 2/CoPt 16/Hf 2/MgO 2/Ta 1.5 (noted as Hf/CoPt 16/Hf). Here, the MgO is a good insulator and the Hf is an amorphous metal that generates no detectable spin current [32] but diminishes any spin-orbit coupling effects at the interfaces (see Refs. [33-35] for the removal of two-magnon scattering and spin memory loss). As characterized in detail in our previous reports [36,37], the CoPt layers are A1-phased polycrystalline films with good composition homogeneity, sharp unoxidized interfaces, and in-plane magnetic anisotropy field of 0.45 T set by the shape anisotropy. The samples were patterned by photolithography and ion milling into 5×60 μm² Hall bars, followed by deposition of 5 nm Ti and 150 nm Pt as electrodes for magnetoresistance measurements. All the experiments in this work are performed at 300 K and under 3 T magnetic field unless otherwise mentioned.

As we show in Figs. 1(b)-1(g), the SiO$_2$/CoPt $t_{CoPt}$/MgO samples exhibit magnetoresistance $\Delta\rho/\rho_0$ ($\Delta\rho \equiv \rho - \rho_0$) with a magnitude of the order of $10^{-3}$ in each of the angle scans ($\alpha$, $\beta$, or $\gamma$), which is giant compared to that of Pt/YIG samples in the literature (typically < $2\times10^{-4}$ for SMR, magnetic-proximity MR, and Hanle MR). The dependences of the MR on the magnetization angles $\alpha$, $\beta$, and $\gamma$ can be fit very well by Eq. (1) (Fig. 1(b)). For each angle scan, the magnetoresistance has a sizable second-order $\cos^4$ contribution in addition to the first-order $\cos^2$ one (Figs. 1(c)-1(g)). As shown in Fig. 1(h), the magnitude of the sum MR, $(\Delta\rho_1+\Delta\rho_2)/\rho_0$, increases for $\alpha$ and $\gamma$ scans but decreases for the $\beta$ scan as the CoPt thickness increases. The latter implies an interface origin of the UMR in the $\beta$ scan. The interface origin of the magnetoresistance is reaffirmed by their being sensitive to the interface details. As shown in Fig. 2(a)-2(c), $(\Delta\rho_1+\Delta\rho_2)/\rho_0$ for $\beta$ scan increases from $1.2\times10^{-3}$ in SiO$_2$/CoPt 16/MgO to $1.7\times10^{-3}$ in the MgO/CoPt 16/MgO and $2.7\times10^{-3}$ in Hf 2/CoPt 16/Hf 2 (more than 2 times greater in magnitude than that of SiO$_2$/CoPt 16/MgO). As shown in Fig. 1(f) and Fig. 2(b), the sum of the $\alpha$-type and $\gamma$-type UMRs always coincides with the $\beta$-type UMR in both magnitude and sign, in excellent agreement with the "sum rule" of the two-vector UMR in Eq. (2). These characteristics (i.e., the interface origin, the occurrence in magnetic single layers, the presence of the $\cos^4$ contributions, and the universal validation of the sum rule) consistently agree with the physics origin of two-vector magnetoresistance for the UMR in the single-layer magnetic metals.

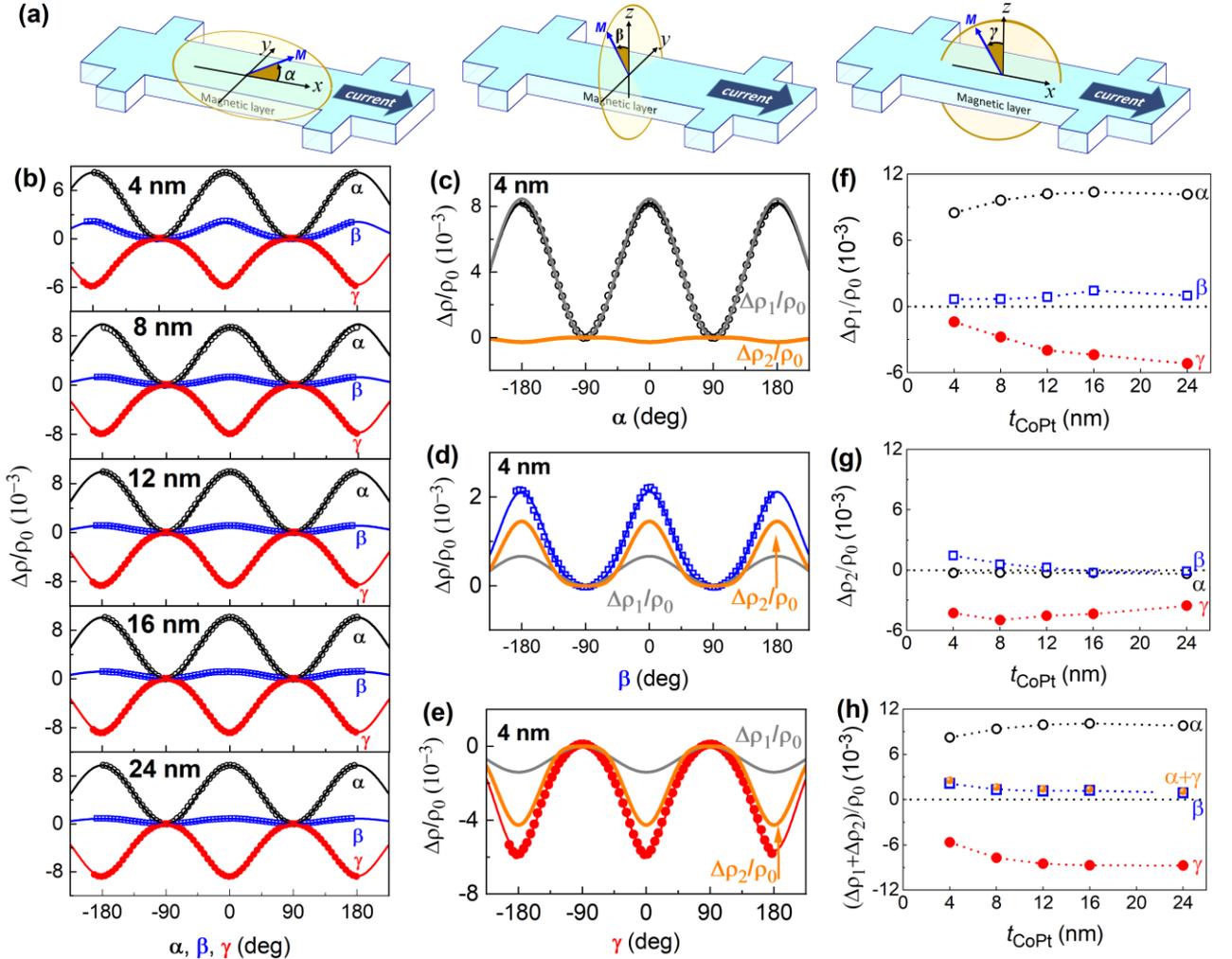

**Fig. 1**. Giant magnetoresistance. (a) Definition of the coordinates and the $\alpha$, $\beta$, and $\gamma$ angles. (b) Dependence on $\alpha$ (black), $\beta$ (blue), and $\gamma$ (red) of the magnetoresistance ($\Delta\rho/\rho_0$) of the SiO$_2$/CoPt/MgO with the CoPt layer thickness of 4 nm, 8 nm, 12 nm, 16 nm, and 24 nm. The solid curves represent the best fits of the data to Eq. (2). (c) $\Delta\rho/\rho_0$ vs $\alpha$, (d) $\Delta\rho/\rho_0$ vs $\beta$, (e) $\Delta\rho/\rho_0$ vs $\gamma$ for the 4 nm CoPt. The gray and orange curves plot the first-order ($\Delta\rho_1\cos^2$) and second-order magnetoresistance contributions ($\Delta\rho_2\cos^4$). Dependences on the CoPt thickness of the magnetoresistance magnitudes (f) $\Delta\rho_1/\rho_0$ and (g) $\Delta\rho_2/\rho_0$. In (f) -(h), the orange dots (marked as $\alpha+\gamma$) are the sum of the $\alpha$-type (black circles) and $\gamma$-type MRs (red dots) and coincide well with the $\beta$-type MR (blue squares). The error bars of the points are smaller than the data symbols.

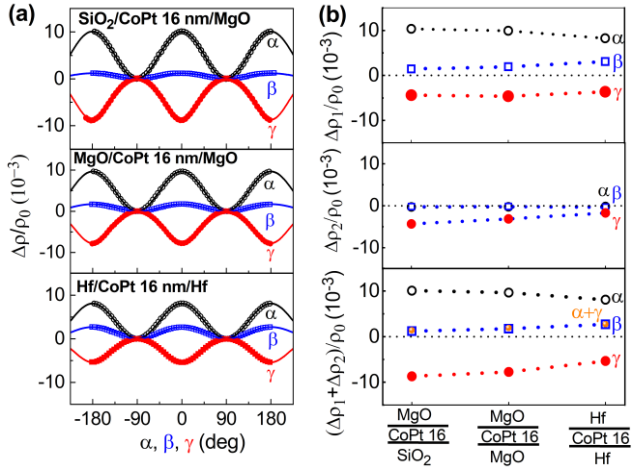

**Fig. 2.** Interface effect. (a) Dependence on $\alpha$ (black), $\beta$ (blue), and $\gamma$ (red) of the magnetoresistance ($\Delta\rho/\rho_0$), (b) the first-order contribution $\Delta\rho_1/\rho_0$, and (c) the second-order contribution $\Delta\rho_2/\rho_0$ for the SiO$_2$/CoPt 16/MgO, MgO/CoPt 16/MgO, and Hf/CoPt 16/Hf. The solid curves in (a) represent the best fits of the data to Eq. (1). In (b), the orange dots (marked as $\alpha+\gamma$) are the sum of the $\alpha$-type MR (black circles) and the $\gamma$-type MR (red dots) and coincide well with the $\beta$-type MR (blue squares). The error bars of the points are smaller than the data symbols.

To obtain a unified understanding of the physics origin of the UMRs of different magnetic heterostructures, we demonstrate below that when revisited, the literature data that were used to claim SMR and other MRs either spin-current related or unrelated can be understood well by the two-vector MR theory. First, the literature UMR data typically include a nonnegligible or even dominating $\cos^4\beta$ contribution that was usually overlooked in the literature and led to significant deviation from a $\cos^2\beta$ dependence in the literature plots. We have plotted the representative literature data of $\beta$-dependent UMRs and their $\cos^2\beta$ (gray curve) and $\cos^4\beta$ (orange curve) components in Fig. 3(a). The $\cos^4\beta$ contribution has also been identified in some reports, such as on MgO/CoFe (termed as "intrinsic AMR")[38,39] and Pt/Ni/Pt (termed as crystalline symmetry-related AMR)[40]. Furthermore, as plotted in Fig. 3(b), the three types of UMRs of all different systems in the literature universally follow the sum rule of Eq. (2) predicted by the two-vector MR theory (i.e., the sum of $\alpha$ and $\gamma$ type MRs equals the $\beta$ type MR under the same angle definition in Fig. 1(a); at least, one of the three MRs equal to the sum of the rest, if the angle definition was different from Fig. 1(a)). More examples that consistently validate the universality of the sum rule are provided in Figs. 3(c) and (d), i.e., the literature data of Pt/YIG (termed as the "SMR" and "hybrid MR" in [5,41]), of Pt/Py/Pt, Au/Py/Au, and SiO$_2$/Py/SiO$_2$ (termed as "hybrid MR" in [5]).

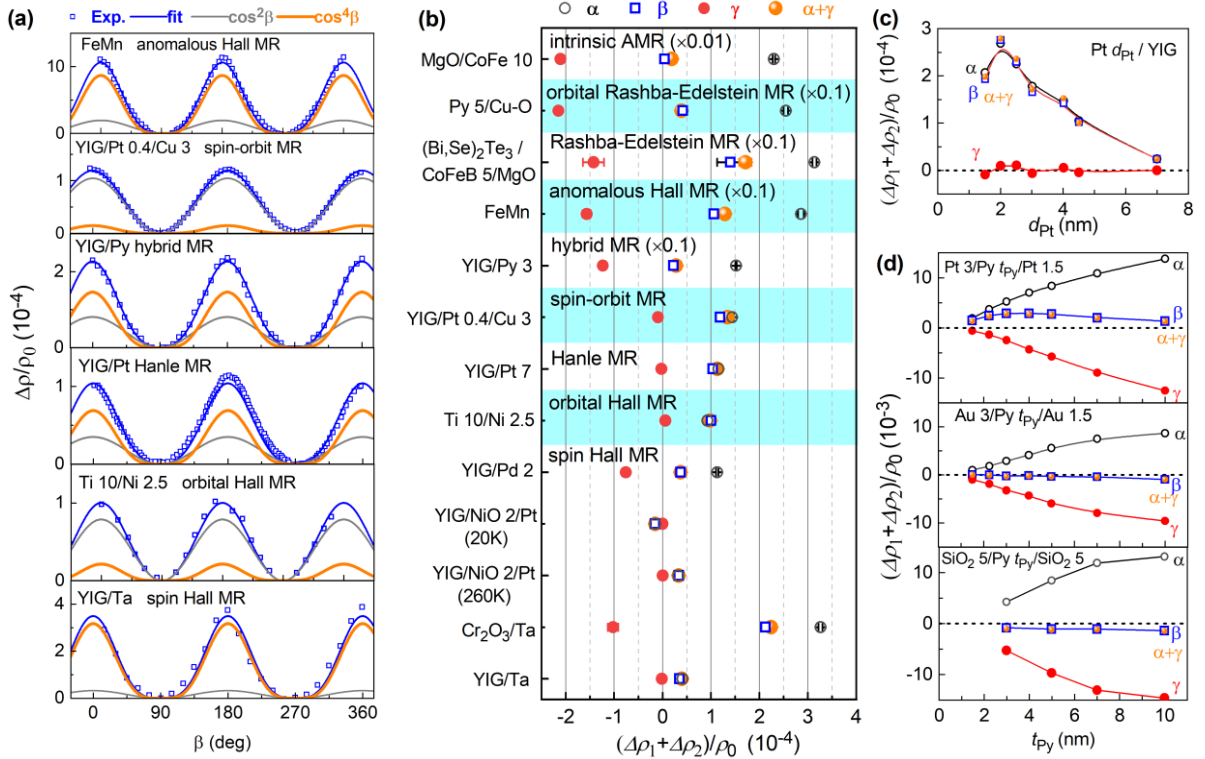

**Fig. 3. High-order UMR and sum rule of the literature data.** (a) Dependences on $\beta$ of UMR for YIG/Ta (termed as "spin Hall MR" in [3]), Ti 10/Ni 2.5 (termed as "orbital Hall MR" in [27]), YIG/Pt (termed as "Hanle MR" in [29]), and YIG/Py (termed as "hybrid MR" in [5]), YIG/Pt 0.4/Cu 3 (termed as "spin-orbit MR" in [25]), and FeMn (termed as "anomalous Hall MR" in [26]), revealing the presence of a large second-order UMR signal. The blue curves plot the best fits of the experimental data to Eq. (1), while the gray and orange curves plot the first-order contribution (with a $\cos^2\beta$ scaling) and the second-order contribution (with a $\cos^4\beta$ scaling). (b) Sum rule of the $\alpha$-, $\beta$-, $\gamma$- type UMRs for YIG/Ta [3], Cr$_2$O$_3$/Ta [11], YIG/NiO 2/Pt (260 K and 20 K)[10], YIG/Pd 2 [6], Ti 10/Ni 2.5 [27], YIG/Pt 7 [29], YIG/Pt 0.4/Cu 3 [25], YIG/Py 3 [5], FeMn [26], (Bi,Se)$_2$Te$_3$/CoFeB 5/MgO [24], Py 5/Cu-O [28], MgO/CoFe 10 [9]. (c) $(\Delta\rho_1 + \Delta\rho_2)/\rho_0$ vs $d_{Pt}$ for Pt $d_{Pt}$/YIG [1,5], (d) $(\Delta\rho_1 + \Delta\rho_2)/\rho_0$ vs $d_{Py}$ for Pt 3/Py $t_{Py}$/Pt 1.5, Au 3/ Py $t_{Py}$/Au 1.5, and (d) SiO$_2$ 5/Py $t_{Py}$/ SiO$_2$ 5 [5]. In (b)-(d), the orange dots (marked as $\alpha+\gamma$) are the sum of the $\alpha$-type MR (black circles) and the $\gamma$-type MR (red dots) and coincide well with the $\beta$-type MR (blue squares), revealing the universal sum rule of Eq. (2).

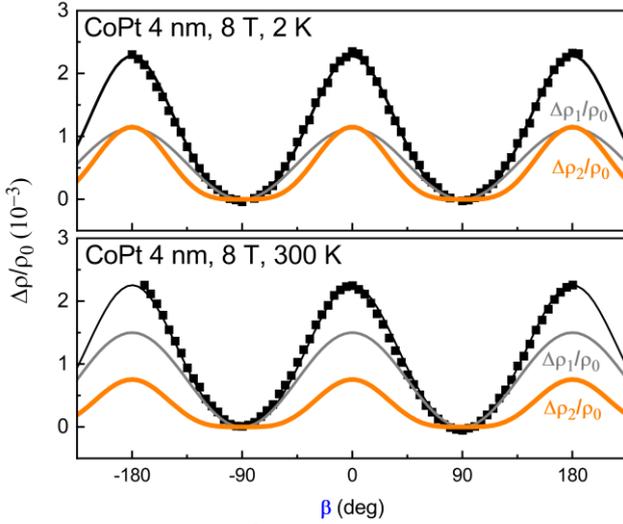

Fig. 4 Dependence on $\beta$ of the unusual magnetoresistance ($\Delta\rho/\rho_0$) for the SiO$_2$/CoPt 4/MgO at temperatures of 2 K and 300 K measured under a magnetic field of 8 T. The black curves represent the best fits of the data to Eq. (2), the gray and orange curves plot the first-order and second-order magnetoresistance contributions, ($\Delta\rho_1/\rho_0$)cos$^2\beta$ and ($\Delta\rho_2/\rho_0$)cos$^4\beta$.

After we have established the good agreement of the two-vector UMR model with the experimental results of the different magnetic single layers, bilayers, and multilayers, we discuss the possible alternative mechanisms. The "intrinsic AMR" model [38] relies on the very specific band structure of the CoFe single crystal and cannot explain universal occurrence of UMR in other systems. SMR, *if we did not question whether its model was theoretically reasonable*, can fulfil some, not all, of the experiments. First, within the SMR frame, the reflection of the spin Hall spin current at the surfaces could generate SMR in a magnetic layer within non-zero SHE, with the contribution of the two surfaces add rather than subtract. However, SMR due to spin-current generation at the magnetic interfaces is less likely as revealed by the absence of spin-current generation at interfaces even with strong interfacial spin-orbit coupling [42]. Second, while the high-order cos$^4\beta$-dependent UMRs ($n\geq 2$) were unexpected in the existing reports of spin-current-related MRs and discussed as inconsistent with SMR in some works [40,43], we find that they could arise within the SMR frame due to the misalignment of the magnetization from the external magnetic field ($H$) when the latter is not much-much greater than the in-plane anisotropy field ($H_k$). For the $\beta$-type SMR at the second order we get

$$\Delta\rho/\rho_0 = \Delta\rho_{1\beta}\left[(1-2H_k/H)\cos^2\beta + (2H_k/H)\cos^4\beta\right]. \quad (3)$$

However, quantitatively, Eq. (3) fails to explain the relative strength of the cos$^4\beta$- and cos$^2\beta$ components of UMR. For example, the relative strength varies significantly rather than remains constant ($\Delta\rho_{2\beta}/\Delta\rho_{1\beta} = 0.43$) for the CoPt samples with $\mu_0H_k = 0.45$ T and $\mu_0H = 3$ T in Figs. 1(f) and 1(g). The discrepancy from SMR is more evident in the cases that the cos$^4\beta$ component of UMR is much greater than the cos$^2\beta$ component for magnetic systems with small $\mu_0H_k$ under very strong magnetic field of several Teslas, e.g., 4 nm CoPt in Fig. 1(d), YIG/HM and FeMn in Fig. 3(a), and Pt/[Co/Ni]$_n$ [43]. As shown in Fig. 4, we have also tested that the cos$^4\beta$ component for the 4 nm CoPt remains significant at 300 K and even greater at low temperatures than the cos$^2\beta$ component when the external field is increased to 8 T, at which Eq. (3) only expect a negligible cos$^4\beta$ component ($\Delta\rho_{2\beta}/\Delta\rho_{1\beta} = 0.1$).

There have also been a number of other experimental indications against SMR. UMR in metallic magnet systems ranges typically from 10$^{-3}$-10$^{-2}$ (see Fig. 1(h), Fig.2(b), and Figs. 3(b)-3(d)), which appears to be too large to be accounted by spin-current effects. The $\beta$-type UMR is strongly dependent on the magnetic layer (e.g., very large in W/Co bilayer but reduced in W/CoFeB bilayer)[23] and increases with the FM thickness to unreasonably large values (e.g., in Pt/Co and Pt/[Co/Ni]$_n$)[23,43], which is in sharp contrast to the expectation that SMR should be independent of the type of the FM layer (the interfacial spin-mixing conductance of a metallic magnet interface is robust against the type of the FM [33] and magnetization [44]) and decrease as the FM thickness increases. More strikingly, the negative sign of the $\beta$-type UMR in Au/Py/Au and SiO$_2$/Py/SiO$_2$ in Fig. 3(d) disagrees with the spin-current-related MR models that the resistivity must be the smallest when the magnetization is in the $y$ direction and thus parallel to the polarization of the spin current (regardless of the sources of the spin current). Therefore, SMR is, if not always absent, not a universal physics origin for UMR.

Theoretically, previous linear response theories [45] have suggested that, when $\vec{\rho}$ was a function of $\vec{m}$ only, $\vec{\rho}(\vec{m})$ would be only allowed to have the conventional anisotropic magnetoresistance with the form $\rho_{ij} = a\delta_{ij} + b\varepsilon_{ijk}m_k + cm_im_j$ ($a$, $b$, and $c$ are coefficients depending possibly only on other scalar parameters such as temperature and disorder configuration other than the direction of $\vec{m}$, $m_i$, $m_j$, and $m_k$ are the three components of $\vec{m}$, $\delta_{ij}$ is Kronecker symbol, $\varepsilon_{ijk}$ is the Levi-Civita symbol following the Einstein summation convention) but cannot include terms like $am_y^2 + b(m_x^2 + m_z^2)$ with $a \neq b$. The two-vector UMR model [31] has further verified that when both the vectors $\vec{m}$ and $\vec{n}$ enter the resistivity tensor, $\vec{\rho}(\vec{m},\vec{n})$ can have nonzero $m_z^2$ term (i.e., the $\beta$-type UMR). If the linear response theories and two-vector UMR theory are correct, the two-vector UMR would be distinct from and more accurate than SMR and other spin-current-based MRs that assume $\vec{m}$ as the only macroscopic vector of the resistivity tensor [1-4,23-30].

In summary, we have presented the universal UMR, including its interface origin, occurrence in magnetic single layers, presence or even dominance of the high-order UMR contributions (e.g., cos$^4$), and universal validity of the sum rule. These results consistently reveal the beautiful agreement of UMR with the physics origin of two-vector MR. Revisiting the literature data reveals that the data that were used to claim SMR and other MRs related or unrelated to spin current can be understood well by the two-vector MR theory, without involving any spin current effect. Experimental and theoretical results against the spin-current-related MRs, but not the two-vector MR, are also extensively discussed. This work presents the first experimental validation of the two-vector MR theory. We believe our results will stimulate efforts towards a unified, precise understanding of the universal UMR phenomena in various spintronic heterostructures.

The authors thank Daniel C Ralph for the fruitful discussions and Changmin Xiong for help with the PPMS measurements. This work was supported partly by the Beijing Natural Science Foundation (Z230006), by the National Key Research and Development Program of China (2022YFA1204000), and the National Natural Science Foundation of China (12274405, 12304155, 12374122). X.R.W. also acknowledge support from the National Key Research and Development Program of China (2020YFA0309600), University Development Fund of the Chinese University of Hong Kong, Shenzhen, and Hong Kong Research Grants Council Grants (16302321, 16300522, and 16300523).